\documentclass{aa}  
\usepackage{epstopdf}
\usepackage{epsfig}
\usepackage{graphicx}
\usepackage{natbib}
\usepackage{gensymb}
\usepackage{txfonts}
\usepackage{amsmath}
\usepackage{dblfloatfix}
\usepackage{caption}
\usepackage[section]{placeins}
\usepackage{hyperref}
\usepackage{float}

\begin{document} 

   \title{CARMENES input catalog of M~dwarfs}

   \subtitle{VI. A time-resolved Ca~{\sc ii}~H\&K catalog from archival data}
   
   \titlerunning{Time-resolved Ca~{\sc ii}~H\&K catalog of CARMENES M dwarfs}

   \author{V.~Perdelwitz
          \inst{1,2}
          \and
          M.~Mittag\inst{2}
          \and
          L.~Tal-Or\inst{1,3}
          \and
          J.\,H.\,M.\,M.~Schmitt\inst{2}
          \and
          J.\,A.~Caballero\inst{4}
          \and
          S.\,V.~Jeffers\inst{5}
          \and
          A.~Reiners\inst{3}
          \and
          A.~Schweitzer\inst{2}
          \and
          T.~Trifonov\inst{6}
          \and
          I.~Ribas\inst{7,8}
          \and
          A.~Quirrenbach\inst{9}
          \and
          P.\,J.~Amado\inst{10}
          \and
          W.~Seifert\inst{9}
          \and
          C.~Cifuentes\inst{4}
          \and
          M.~Cort\'es-Contreras\inst{4}
          \and
          D.~Montes\inst{11}
          \and
          D.~Revilla\inst{11}
          \and
          S.\,L.~Skrzypinski\inst{11}
          }

   \institute{
             Department of Physics, Ariel University, Ariel 40700, Israel\\
             \email{volkerp@ariel.ac.il}
         \and
             Hamburger Sternwarte, Universit\"at Hamburg, Gojenbergsweg 112, 21029 Hamburg, Germany
         \and
             Institut f\"ur Astrophysik, Georg-August-Universit\"at, Friedrich-Hund-Platz 1, 37077 G\"ottingen, Germany
         \and
             Centro de Astrobiolog\'ia (CSIC-INTA), ESAC, Camino bajo del Castillo s/n, 28692 Villanueva de la Cañada, Madrid, Spain
         \and 
             Max-Planck-Institut f\"ur Sonnensystemforschung, Justus-von-Liebig-Weg 3, 37077 Göttingen
         \and
             Max-Planck-Institut f\"ur Astronomie, K\"onigstuhl 17, 69117 Heidelberg, Germany
         \and
             Institut de Ci\`encies de l’Espai (ICE, CSIC), Campus UAB, C/Can Magrans s/n, 08193 Bellaterra, Spain
         \and
             Institut d’Estudis Espacials de Catalunya (IEEC), 08034 Barcelona, Spain
         \and
             Landessternwarte, Zentrum für Astronomie der Universit\"at Heidelberg, K\"onigstuhl 12, 69117 Heidelberg, Germany
         \and
             Instituto de Astrof\'isica de Andaluc\'ia (IAA, CSIC), Glorieta de la Astronom\'ia 1, 18008 Granada, Spain
         \and
             Departamento de F\'isica de la Tierra y Astrof\'isica and IPARCOS-UCM (Instituto de F\'isica de Part\'iculas y del Cosmos de la UCM), Facultad de Ciencias F\'isicas, Universidad Complutense de Madrid, 28040, Madrid, Spain
             }

   \date{Received 25 March 2021 / Accepted 06 July 2021}
 
  \abstract 
   {Radial-velocity (RV) jitter caused by stellar magnetic activity is an important factor in state-of-the-art exoplanet discovery surveys such as CARMENES. 
   Stellar rotation, along with heterogeneities in the photosphere and chromosphere caused by activity, can result in false-positive planet detections. Hence, it is necessary to determine the stellar rotation period and compare it to any putative planetary RV signature. 
   Long-term measurements of activity indicators such as the chromospheric emission in the Ca~{\sc ii}~H\&K lines ($R_{\mathrm{HK}}^\prime$) enable the identification of magnetic activity cycles.}
   {In order to determine stellar rotation periods and study the long-term behavior of magnetic activity of the CARMENES guaranteed time observations (GTO) sample, it is advantageous to extract $R_{\mathrm{HK}}^\prime$ time series from archival data, since the CARMENES spectrograph does not cover the blue range of the stellar spectrum containing the Ca~{\sc ii}~H\&K lines.}
   {We have assembled a catalog of 11\,634 archival spectra of 186 M dwarfs acquired by seven different instruments covering the Ca~{\sc ii}~H\&K regime: ESPADONS, FEROS, HARPS, HIRES, NARVAL, TIGRE, and UVES. The relative chromospheric flux in these lines, $R_{\mathrm{HK}}^\prime$, was directly extracted from the spectra by rectification with PHOENIX synthetic spectra via narrow passbands around the Ca~{\sc ii}~H\&K line cores.}
   {The combination of archival spectra from various instruments results in time series for 186 stars from the CARMENES GTO sample. As an example of the use of the catalog, we report the tentative discovery of three previously unknown activity cycles of M dwarfs.}
   {We conclude that the method of extracting $R_{\mathrm{HK}}^\prime$ with the use of model spectra yields consistent results for different instruments and that the compilation of this catalog will enable the analysis of long-term activity time series for a large number of M dwarfs.}

   \keywords{stars: activity --
            planetary systems --
            techniques: radial velocities --
            stars: late-type
               }

   \maketitle


\section{Introduction}
In recent years, low-mass stars have become the focus of the search for exoplanets via radial-velocity (RV) signals. The fact that the habitable zones of M~dwarfs are much closer to the host star than those of solar analogs, combined with their low masses and radii, means that the detection of Earth-mass planets in a temperate environment is comparatively easy. There have been several exoplanet search campaigns targeting low-mass stars \citep[e.g.,][]{2009A&A...505..859Z,2013A&A...549A.109B,2013ApJ...775...91B}, contributing to a total of 234 known exoplanets orbiting M dwarf hosts to date\footnote{Based on the NASA Exoplanet Archive  as of March 2021 (\url{http://exoplanetarchive.ipac.caltech.edu}).}. Several stars in the immediate neighborhood of the Sun have recently been found to harbor Earth-sized exoplanets within their habitable zone, such as our closest neighbor \object{Proxima Centauri} \citep{2016Natur.536..437A} and the close-by \object{Barnard's Star} \citep{2018Natur.563..365R}, \object{Teegarden's Star} \citep{2019A&A...627A..49Z}, and \object{Lalande~21185} \citep[][and references therein]{2020A&A...643A.112S}.

One of the surveys dedicated to discovering planets orbiting M~dwarfs is CARMENES \citep{2010SPIE.7735E..13Q,2018A&A...612A..49R}, an instrument built for the 3.5\,m telescope at the Calar Alto Observatory. 
Being designed for high-precision RV measurements of cool stars, it comprises two spectrograph channels with wavelength ranges of 0.52 to 0.96\,$\mu$m and 0.96 to 1.71\,$\mu$m, respectively, with spectral resolutions of $\mathcal{R} \approx$ 80\,000--96\,000 \citep{2014SPIE.9147E..1FQ} capable of providing an RV accuracy of $\sim 1.2$\,m\,s$^{-1}$ \citep{2020A&A...640A..50B}. More than 350 stars from the CARMENES guaranteed time observations (GTO) sample have been monitored since January 2016, resulting in 26 planet detections to date \citep[e.g.,][see \citealt{2021arXiv210703802S} for an exhaustive review]{2020A&A...639A.132B,2020A&A...642A.173N}, 
as well as several planet confirmations \citep{2018A&A...609A.117T,2020A&A...638A..16T,2018AJ....155..257S,2019A&A...627A.116L,2020A&A...636A.119S}.

The study of stellar magnetic activity plays a crucial role in the search for Earth-like planets around M~dwarfs. 
These stars can be very active, resulting in a high level of irradiation in the UV and X-ray regimes for planets within their habitable zones \citep[e.g.,][]{2016PhR...663....1S,2017ApJ...841..124V,2017ApJ...843...31Y,2019AsBio..19...64T}. 
Regarding the detection of planets via RV modulations, activity can pose a problem, as the stellar rotation, along with heterogeneities in the photosphere and chromosphere (that is, spots, plages, etc.), can induce a radial velocity signal of several m\,s$^{-1}$, thus masking or mimicking planetary signals \citep{2010A&A...520A..53L,2018A&A...614A.122T,2020A&A...641A..69B}. In order to be able to detect RV modulations caused by companions with small semi-amplitudes, it is thus necessary to understand stellar activity and its influence on the stellar RV signal \citep{2007A&A...473..983D,2010ApJ...710..432R,2013A&A...557A..93F} and develop methods to mitigate this effect \citep{2014ApJ...796..132D,2015MNRAS.452.2269R,2020A&A...636A..36L}.

The emission in line cores of several atomic transitions can be used to trace magnetic activity in different layers of the chromosphere \citep{1981ApJS...45..635V,Hall2008}.
The most prominent of these chromospheric indicators are the Ca~{\sc ii}~H\&K $\lambda\lambda$3933.660,3968.47\,{\AA} doublet \citep{1968ApJ...153..221W}, the Ca {\sc ii} infrared triplet \citep[IRT; ][]{1937CMWCI.575....1W,1972SoPh...25..357S,2017A&A...607A..87M}, and the H$\alpha$ line \citep{1986ApJS...60..551F}. 

A common approach to deriving the chromospheric flux from spectroscopic data is the subtraction of appropriate model atmospheres, or comparison spectra of inactive stars with similar parameters, from measured spectra. 
Named the ``spectral subtraction technique'' by \cite{1995A&AS..114..287M}, variations of it had been in use around a decade earlier \citep{1985ApJ...295..162B,1985ApJ...289..269H}. Since then, it has been employed by various authors to measure the chromospheric flux excess in the Ca~{\sc ii}~H\&K lines, for example, for FGK-type stars using PHOENIX model atmospheres by \cite{2013A&A...549A.117M}, or for M dwarfs using BT-Settl models \citep{2014IAUS..299..271A} by \cite{2017A&A...598A..28S}. 
Other authors have employed different techniques to determine the chromospheric emission in Ca~{\sc ii}~H\&K for M dwarfs, such as
\citet{2015MNRAS.452.2745S, 2018A&A...612A..89S} or \citet{2017A&A...600A..13A}, who calibrated and converted the $S$~index \citep{1968ApJ...153..221W} extracted from HARPS spectra.

\begin{table}[]
\centering
\small
\caption{Source instruments for the database.} \label{table:2}
\begin{tabular}{lcccc}       
\hline\hline               
Instrument & $\mathcal{R}$ [$10^3$] & References & Archive links & \# \\ \hline
\noalign{\smallskip}
ESPADONS & 80 & 1, 2 & 10 & 577\\ 
FEROS & 48 & 3 & 12 & 345\\
HARPS & 115 & 4 & 12 & 4057\\
HIRES & 25--85 & 5, 6 & 12 & 4232\\
NARVAL & 65-75 & 2, 7 & 10 & 2283\\
TIGRE & 20 & 8 & 13 & 131\\
UVES & 22--78 & 9&11 &68\\ 
\noalign{\smallskip}
\hline
\noalign{\smallskip}
{Total} & & & & {11\,634}\\ 
\noalign{\smallskip}
\hline
\end{tabular}
\tablebib{
(1)~\citet{2006ASPC..358..362D};
(2) \citet{2014PASP..126..469P}; 
(3) \citet{1999Msngr..95....8K}; 
(4) \citet{2003Msngr.114...20M};
(5) \citet{1994SPIE.2198..362V}; 
(6) \citet{2002ASPC..270....5V}; 
(7) \citet{1997MNRAS.291..658D};
(8) \citet{2014AN....335..787S};
(9) \citet{2000SPIE.4008..534D}; 
(10) \url{http://polarbase.irap.omp.eu/};
(11) \url{http://archive.eso.org/scienceportal/home};
(12) \url{koa.ipac.caltech.edu/};
(13) \url{https://hsweb.hs.uni-hamburg.de/projects/TIGRE/EN/hrt_user/hrt_user_main_page.html}.
}
\end{table}

The study of long-term magnetic cycles and their possible relation to stellar rotation periods may provide further insight into the dynamo mechanism \citep{1996ApJ...460..848B,2007ApJ...657..486B,2015MNRAS.449.3471S}. 
Several such cycles in M~dwarfs have been detected using photometric and spectroscopic data \citep[][and references therein]{2012ARep...56..716S,2013ApJ...764....3R,2016A&A...595A..12S,2019A&A...621A.126D}.
Chromospheric activity indices for the CARMENES sample of M~dwarfs have been studied by various authors \citep{2018A&A...612A..49R,2019A&A...623A..44S}. 
Besides, \cite{2018A&A...614A..76J} studied the sensitivity of, and correlations between, the indicators within the CARMENES spectral range with a larger sample, while \cite{2019A&A...623A..24F}  used time series of H$\alpha$, Na~{\sc ii} D$_1$ and D$_2$, and Ca~{\sc ii}~IRT from CARMENES spectra to infer previously unknown stellar rotation periods. 
Since the CARMENES spectrograph does not cover the Ca~{\sc ii}~H\&K regime, \cite{2018A&A...614A..76J} used FEROS, CAFE \citep{2013A&A...552A..31A} and HRS \citep{1998SPIE.3355..387T} spectra to obtain emission levels in these lines for a large number of targets from the CARMENES GTO sample.

The aim of this work is to use a variant of the spectral subtraction technique and assemble a database of archival spectra of the CARMENES GTO sample acquired with seven instruments, from which $R_{\mathrm{HK}}^\prime$, that is, the chromospheric emission in Ca~{\sc ii}~H\&K relative to the bolometric flux, is extracted in a homogeneous fashion. 
The main purpose of this catalog of $R_{\mathrm{HK}}^\prime$ time series is to enable the determination of stellar rotation periods. 
The publication is structured in the following manner. In Sect.~\ref{sec:2} we give an overview of the used data, the reduction of which is then described in Sect.~\ref{sec:3}. 
Next, Sect.~\ref{sec:4} presents the results, including a comparison to previous publications, and, as an example of applications, the long-term behavior of three targets with sufficient coverage, followed by the conclusions in Sect.~\ref{sec:5}.

\section{Data acquisition and analysis}
\label{sec:2}

   \begin{figure*}[]
   \centering
   \includegraphics[width=1.00\textwidth]{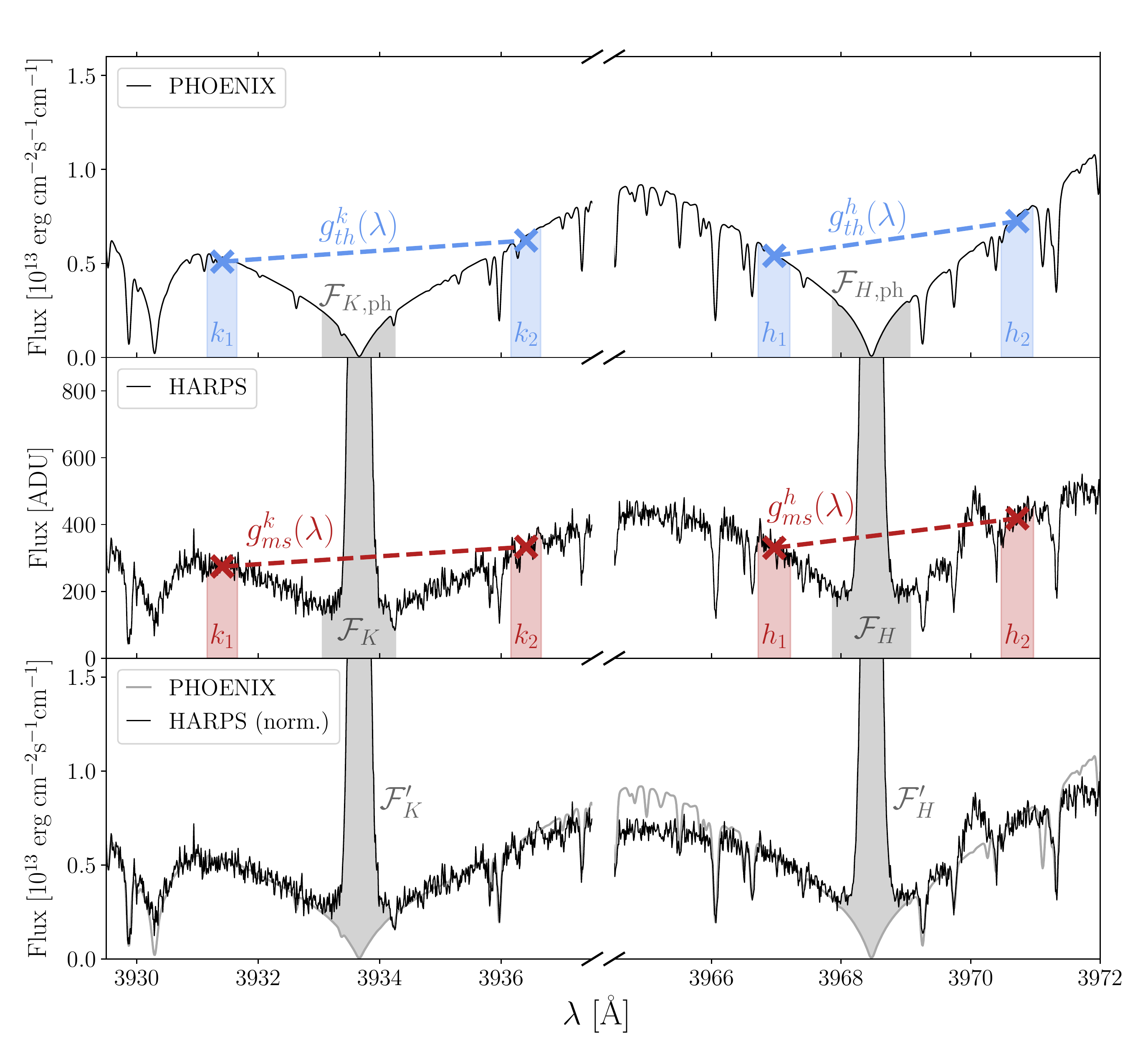}
      \caption{Schematic of the direct determination of $R_{\mathrm{HK}}^\prime$ from a single HARPS spectrum of the M1.5\,V star \object{HD~36395} (Karmn J05314$-$036).
      Its tabulated astrophysical parameters are $T_{\mathrm{eff}}= 3891\pm51$\,K, $\log{g}=4.64\pm0.07$, $\mathrm{[Fe/H]}=0.23\pm0.16$ \citep{2019A&A...625A..68S} and an upper limit of $v \sin{i}$ = 2\,km\,s$^{-1}$ \citep{2018A&A...612A..49R}.
      {\em Top panel}: a PHOENIX synthetic spectrum with parameters $T_{\rm eff}$ = 3900\,K, $\log{g}$ = 4.5, [Fe/H] = 0.0. 
      {\em Middle panel}: the measured HARPS spectrum. 
      The six bands used for rectification of the spectrum and subtraction of the photospheric flux ($H$, $K$, $h_1$, $h_2$, $k_1$, and $k_2$) are marked as shaded areas.
      {\em Bottom panel}: flux-calibrated spectrum (black line) after rectification of the measured spectrum with the gradient functions $g^{h/k}_{th/ms}(\lambda)$ and, for comparison, the PHOENIX spectrum (gray line). 
      The chromospheric fluxes $\mathcal{F_{H/K}^\prime}$ (gray areas in bottom panel) are derived by subtracting the theoretical photospheric fluxes $\mathcal{F}_{\mathrm{H/K}\mathrm{, phot}}$ (gray areas in upper panel) from the surface fluxes $\mathcal{F}_{\mathrm{H/K}}$ (gray areas in middle panel).}
    \label{fig:method}
    \end{figure*}

In order to obtain a catalog of $R_{\mathrm{HK}}^\prime$ for the CARMENES GTO sample, we assembled a database of spectra from seven instruments that cover the spectral region around Ca~{\sc ii}~H\&K. All data were downloaded from the respective archive web pages as pipeline-reduced spectra, either in single-order or order-merged format.
Table~\ref{table:2} gives an overview of the instruments and the number of available spectra of stars in the GTO sample with a signal-to-noise ratio of S/N~$>$~2 as defined in Section~\ref{sec:3}. The resulting database is comprised of a total of 11\,634 spectra of 186 targets, all with a spectral resolution of 20\,000 or higher.

\subsection{Extraction of $R_{\mathrm{HK}}^\prime$}
\label{sec:3}

In this section we give a description of our approach to derive $R_{\mathrm{HK}}^\prime$. Figure~\ref{fig:method}, which shows the Ca~{\sc ii}~H\&K lines in PHOENIX model spectra and HARPS observations, gives an overview of the approach. 
We denote $R_{\mathrm{HK}}^\prime$ as the ratio between Ca~{\sc ii}~H\&K emission and the stellar bolometric flux \citep{1978ApJ...220..619L,1979ApJS...41...47L}:

\begin{equation}
 R_{\mathrm{HK}}^\prime=\frac{\mathcal{F}_{\mathrm{HK}}^\prime}{\mathcal{F}_{\mathrm{bol}}}=\frac{\mathcal{F}_{\mathrm{H}}^\prime+\mathcal{F}_{\mathrm{K}}^\prime}{\sigma T_{\mathrm{eff}}^4},
\end{equation}

\noindent where $T_{\mathrm{eff}}$ is the stellar effective temperature, $\sigma$ is the Stefan-Boltzmann constant, and $\mathcal{F}_{\mathrm{H}}^\prime$ and $\mathcal{F}_{\mathrm{K}}^\prime$ are the chromospheric fluxes:

\begin{equation}
\begin{split}
 \mathcal{F}_{\mathrm{H}}^\prime=\mathcal{F}_{\mathrm{H}}-\mathcal{F}_{\mathrm{H}\mathrm{, phot}},\\
\mathcal{F}_{\mathrm{K}}^\prime=\mathcal{F}_{\mathrm{K}}-\mathcal{F}_{\mathrm{K}\mathrm{, phot}},
\end{split}
\end{equation}

\noindent that is, the difference between the surface fluxes $\mathcal{F}_{\mathrm{H}}$ and $\mathcal{F}_{\mathrm{K}}$ and the photospheric fluxes $\mathcal{F}_{\mathrm{H}\mathrm{, phot}}$ and $\mathcal{F}_{\mathrm{K}\mathrm{, phot}}$.

The oldest approach to derive $R_{\mathrm{HK}}^\prime$ has been to determine the Mount Wilson $S$~index \citep{1968ApJ...153..221W,1991ApJS...76..383D} and subsequently transform it via color-dependent conversion factors \citep{1982A&A...107...31M,1984ApJ...279..763N}: 

\begin{equation}
 R_{\mathrm{HK}}^\prime=C_{\rm cf}~ S - R_{\mathrm{phot}},
\end{equation}

\noindent in which the conversion factor $C_{\rm cf}$ and the ratio between photometric and bolometric flux $R_{\mathrm{phot}}=(\mathcal{F}_{\mathrm{H}\mathrm{, phot}}+\mathcal{F}_{\mathrm{K}\mathrm{, phot}})/\mathcal{F}_{\mathrm{bol}}$ are color-dependent \citep[e.g.,][]{2017A&A...600A..13A}.

Since our database contains archival spectra from seven different instruments with spectral resolutions $\mathcal{R}$ ranging from $\approx2\,10^4$ to $\geq10^5$ and we wanted to avoid any transformation via the $S$~index, we chose to determine the relative flux excess with a variant of the spectral subtraction technique using a set of narrow wavelength bands around the H and K lines to rectify the spectra with PHOENIX synthetic spectra \citep[][ \url{http://phoenix.astro.physik.uni-goettingen.de/}]{2013A&A...553A...6H}.
The approach can be summarized in the following steps:

\begin{table}[]
\centering
\small
\caption{Wavelength bands used for rectification and flux extraction$^{a}$.} 
\label{table:1}
\begin{tabular}{cc}       
\hline\hline
\noalign{\smallskip}
Band & Wavelength range [$\lambda_{\rm min}$, $\lambda_{\rm max}$] \\ 
\noalign{\smallskip}
\hline
\noalign{\smallskip}
$K$ & [$\lambda_{K}-0.6$~\AA, $\lambda_{K}+0.6$~\AA] \\ 
$H$ & [$\lambda_{H}-0.6$~\AA, $\lambda_{H}+0.6$~\AA] \\ 
$k_1$ & [$\lambda_{K}-2.5$~\AA, $\lambda_{K}-2.0$~\AA] \\ 
$k_2$ & [$\lambda_{K}+2.5$~\AA, $\lambda_{K}+3.0$~\AA] \\ 
$h_1$ & [$\lambda_{H}-1.75$~\AA, $\lambda_{H}-1.25$~\AA] \\ 
$h_2$ & [$\lambda_{H}+2.0$~\AA, $\lambda_{H}+2.5$~\AA] \\ 
\noalign{\smallskip}
\hline
\end{tabular}
\tablefoot{
\tablefoottext{a}{Relative to the central wavelengths: $\lambda_{K}$ = 3933.66\,{\AA}, $\lambda_{H}$ = 3968.47\,{\AA}.}
}

\end{table}

\begin{itemize}
 \item Each spectrum in the grid of PHOENIX synthetic spectra is rotationally broadened using the {\sc PyAstronomy} package \citep{2019ascl.soft06010C} with rotational velocities in the range of $v \sin{i} = [0,100]$\,km\,s$^{-1}$ in increments of 1\,km\,s$^{-1}$ with a linear limb-darkening law and coefficients from \cite{2011A&A...529A..75C} for the Johnson $U$ filter.
 
 \item Each spectrum from this extended grid is then instrumentally broadened with the appropriate spectral resolution.
 
 \item Mean fluxes in six bands are extracted from each model spectrum, one of width 1.2\,{\AA} around the center of each of the H\&K lines, and in two narrow bands of width 0.5\,{\AA} situated around each line. 
 The passbands, which are listed in Table~\ref{table:1} relative to the central wavelengths and illustrated in Fig.~\ref{fig:method}, were chosen to be {\it (i)} close to the line centers so as to minimize the influence of imperfect blaze normalization, {\it (ii)} in ranges with few absorption lines to mitigate errors in the determination of stellar metallicity, and {\it (iii)} free of emission lines such as $\mathrm{H}\epsilon$ $\lambda$3970.08\,{\AA}.
 This step results in a grid of fluxes for each of the six passbands as a function of the stellar parameters $T_{\mathrm{eff}}$, $\log{g}$, [Fe/H], $v \sin{i}$, and spectral resolution~$\mathcal{R}$.
 
 \item After shifting the pipeline-reduced spectra (Sect.~\ref{sec:2}) to the respective stellar rest frame using velocities from Carmencita \citep[the CARMENES input catalog; ][]{2015A&A...577A.128A,2016csss.confE.148C}, mean fluxes from the same set of passbands (Table~\ref{table:1}) are extracted.
 
 \item For each measured spectrum, the theoretical fluxes are then linearly extrapolated from the grids computed in {step \#3} using the stellar input parameters ($T_{\rm eff}$, $\log{g}$, [Fe/H]) from Carmencita and the spectral resolution of the respective instrument.
 The references for the astrophysical stellar parameters in the catalog, including $v \sin{i}$, are \citet{1998A&A...331..581D}, \citet{2010AJ....139..504B}, \citet{2010A&A...514A..97L}, \citet{2012AJ....143...93R}, \citet{2017AJ....153...21L}, \citet{2018MNRAS.475.1960F}, \citet{2018A&A...614A..76J}, \citet{2018A&A...612A..49R}, and, especially, \citet{2019A&A...625A..68S}. 
 
 \item The gradients $g^h_{th}(\lambda)$, $g^k_{th}(\lambda)$, $g^h_{ms}(\lambda)$, and $g^k_{ms}(\lambda)$ between the sets of passbands $h_1$, $h_2$ and $k_1$, $k_2$ are calculated for the theoretical (``th'') and measured (``ms'') spectrum via:
 
 \begin{equation}
 \label{eq:1}
 \begin{split}
  g^{h/k}_{\rm th/ms}(\lambda) & = \frac{f_{\rm th/ms}(\lambda_{h1/k1})-f_{\rm th/ms}(\lambda_{h2/k2})}{\lambda_{h1/k1}-\lambda_{h2/k2}}\cdot \lambda \\ &+ \frac{f_{\rm th/ms}(\lambda_{h2/k2})\cdot \lambda_{h1/k1}-f_{\rm th/ms}(\lambda_{h1/k1})\cdot \lambda_{h2/k2}}{\lambda_{h1/k1}-\lambda_{h2/k2}}.
 \end{split}
 \end{equation} 
 
\noindent They are plotted in the upper and middle panels of Fig.~\ref{fig:method} with dashed lines.

 \item The measured spectrum can then be rectified in the two wavelength regions around the H\&K lines through:
 
 \begin{equation}
  f^h_{\rm rec}(\lambda)=f^h_{ms}(\lambda) \frac{g^h_{th}(\lambda)}{g^h_{ms}(\lambda)}
 \end{equation}
 
\noindent and:
 
 \begin{equation}
  f^k_{\rm rec}(\lambda)=f^k_{\rm ms}(\lambda) \frac{g^k_{\rm th}(\lambda)}{g^k_{\rm ms}(\lambda)},
 \end{equation}
 
\noindent resulting in normalized, flux-calibrated spectra in the two wavelength regimes, as displayed as a solid black line in the bottom panel of Fig.~\ref{fig:method}.
Assuming that the theoretical model is adequate, this approach will also correct to first order for imperfect blaze normalization of the measured spectra. 

 \item $R_{\mathrm{HK}}^\prime$ is then extracted directly from the rectified spectrum:
 
  \begin{equation}
  \label{eq:5}
  \begin{split}
  R_{\mathrm{HK}}^\prime & = R_{\mathrm{HK}}-R_{\mathrm{phot}}\\
  & = \frac{\mathcal{F}_{H}+\mathcal{F}_{K}}{\sigma T_{\mathrm{eff}}^4}-\frac{\mathcal{F}_{H\mathrm{, ph}}+\mathcal{F}_{K\mathrm{, ph}}}{\sigma T_{\mathrm{eff}}^4}\\
  & =\Bigg(\int_{\lambda_{H}-0.6~\AA}^{\lambda_{H}+0.6~\AA}f^h_{rec}(\lambda)d\lambda+\int_{\lambda_{K}-0.6~\AA}^{\lambda_{K}+0.6~\AA}f^k_{rec}(\lambda)d\lambda\\
  &\;\; -\int_{\lambda_{H}-0.6~\AA}^{\lambda_{H}+0.6~\AA}f_{th}(\lambda)d\lambda -\int_{\lambda_{K}-0.6~\AA}^{\lambda_{K}+0.6~\AA}f_{th}(\lambda)d\lambda\Bigg)\,/\left(\sigma T_{\mathrm{eff}}^4\right).
 \end{split}
 \end{equation} 
 
 \item The error estimation is performed with a Monte Carlo approach. In each trial, the flux in each bin of the measured spectrum is randomly displaced within a gaussian distribution with width of the flux error, and all stellar parameters are displaced in the same manner according to the error values given by Carmencita. $R_{\mathrm{HK}}^\prime$ is then evaluated for each trial, and the error $\Delta R_{\mathrm{HK}}^\prime$ is taken to be the standard deviation of the resulting set.
\end{itemize}

Figure~\ref{fig:method} shows a schematic of the direct determination of $R_{\mathrm{HK}}^\prime$ for a single HARPS spectrum the M1.5\,V star HD~36395. 
The upper panel displays a PHOENIX synthetic spectrum with similar astrophysical parameters broadened with $v \sin{i}=2$\,km\,s$^{-1}$ and the spectral resolution of HARPS ($\mathcal{R}=115\,000$). 
From this theoretical spectrum and the measured HARPS spectrum (middle panel), we extracted fluxes in the six different bands listed in Table~\ref{table:1} (shaded red and blue areas in Fig.~\ref{fig:method})
and, from them, determined the gradient functions $g^{h/k}_{\rm th/ms}(\lambda)$ (dashed lines in upper and middle panel). 
The bottom panel of Fig.~\ref{fig:method} shows the rectified HARPS spectrum, which is in good agreement with the photospheric PHOENIX synthetic spectrum in the areas around the H\&K line cores. 
In order to avoid the extraction of $R_{\mathrm{HK}}^\prime$ from spectra with insufficient flux in the H\&K regime, we required the integrated signal-to-noise ratio of the spectra in the four rectification bands $h_1$, $h_2$, $k_1$, and $k_2$ to be $\left(\mathrm{S/N}\right)_{h_1,h_2,k_1,k_2}\geq1.5$. 
This restriction resulted in a total of 11\,634 $R_{\mathrm{HK}}^\prime$ measurements for 186 targets from the CARMENES GTO sample.
Table~\ref{table:4} shows the first 5 rows of the catalog of 11\,634 $R_{\mathrm{HK}}^\prime$ measurements of 186 M dwarfs.

\subsection{Sensitivity to uncorrected RV shifts and spectral resolution, and comparison between instruments}

   \begin{figure}[]
   \centering
   \includegraphics[width=\hsize]{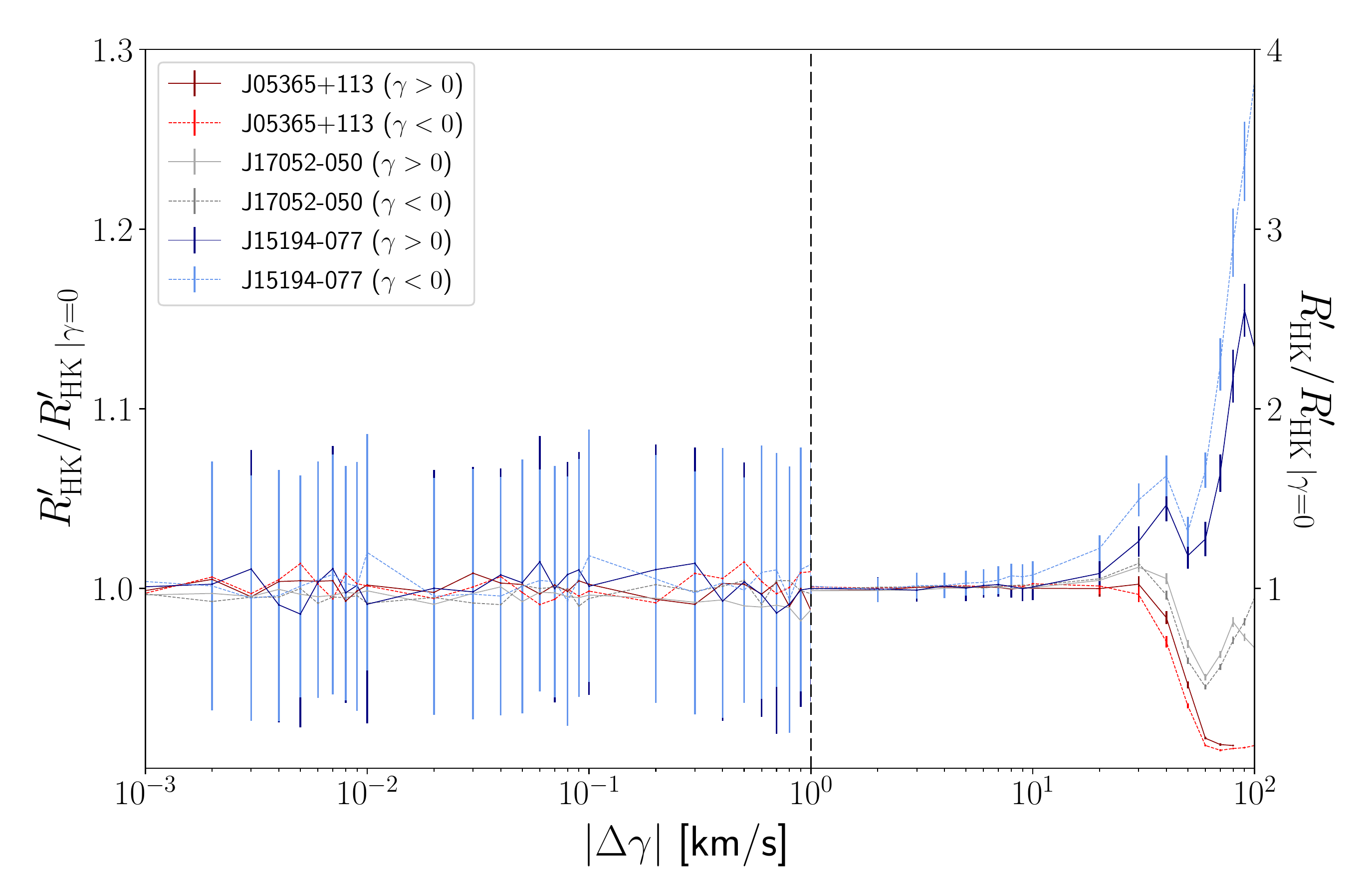}
   \captionsetup{width=\hsize}
      \caption{$R_{\mathrm{HK}}^\prime$ normalized by the rest frame index ${R_{\mathrm{HK}}^\prime}_{|\gamma=0}$ as a function of velocity offset. 
      We note the change of the vertical axis scale to the left and right of $\Delta \gamma \geq1$\,km\,s$^{-1}$ marked by the vertical dashed line.}
         \label{fig:verr}
   \end{figure}
   
Since our algorithm shifts the archival spectra to the stellar rest frame based on the mean radial velocity value listed by Carmencita, and does not account for RV modulations caused by planetary companions or induced by stellar magnetic activity, we tested the stability of the approach with regard to uncorrected RV shifts. 
Single spectra of three early M dwarfs
(\object{V2689~Ori} / Karmn J05365+113, 
\object{HO~Lib} / J15194--077, 
and \object{Wolf~636} / J17052--050) 
of different activity levels (active, semi-active, quiescent) of similar S/N were shifted by a set of radial velocities $\pm j \times 10^i$\,km\,s$^{-1}$ with $j=1:9$ and $i=-3:2$ from their catalog rest frame velocity, after which we extracted $R_{\mathrm{HK}}^\prime$ .
Figure~\ref{fig:verr} shows the resulting $R_{\mathrm{HK}}^\prime$ normalized by the rest frame index ${R_{\mathrm{HK}}^\prime}_{|\gamma=0}$ as a function of the velocity offset, $\Delta \gamma$. 
Since the measurement errors are larger than the deviation caused by velocity offsets $\Delta \gamma\leq10$\,km\,s$^{-1}$, we concluded that the influence of smaller velocity offsets was negligible and that we did not need to correct for the RV of single spectra. 

\begin{figure}[]
\centering
\includegraphics[width=\hsize]{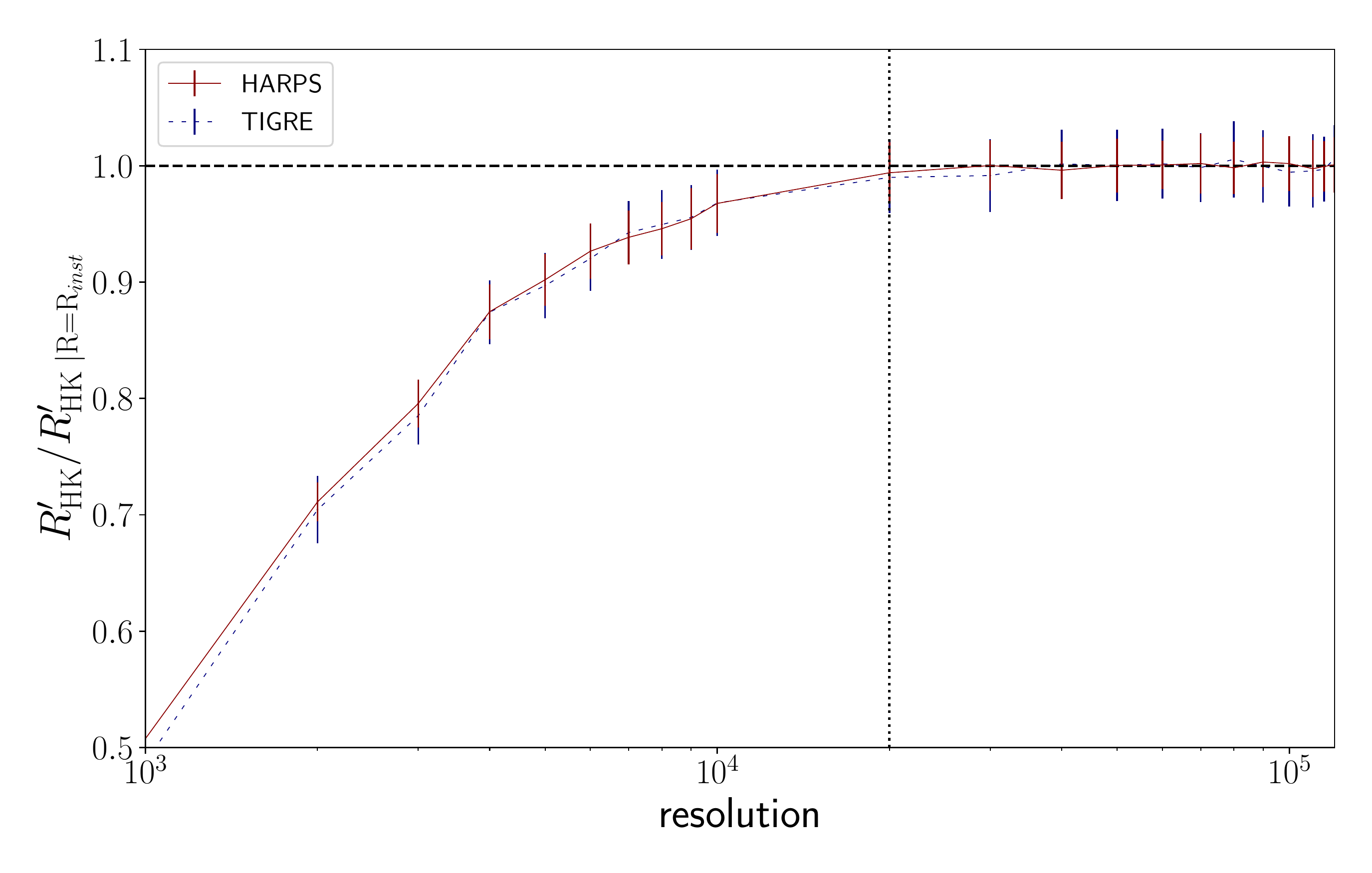}
\captionsetup{width=\hsize}
    \caption{$R_{\mathrm{HK}}^\prime$ normalized by the index ${R_{\mathrm{HK}}^\prime}_{|\mathrm{R}=\mathrm{R}_{inst}}$ as a function of resolution of the template spectrum used during extraction. 
    Both the TIGRE data (dashed blue line) and HARPS data (solid red line) are based on spectra from the same target, \object{HD~119850} / Karmn~J13457+148. 
    Unity is represented by the horizontal dashed line and the resolution of 20\,000 by the dotted vertical line.}
        \label{fig:reserr}
\end{figure}

In a similar manner, we tested whether the artificial broadening with regard to spectral resolution applied in our approach leads to systematic offsets in $R_{\mathrm{HK}}^\prime$, since the model spectra are broadened according to $\mathcal{R}$ and not using the actual line spread function. Figure~\ref{fig:reserr} shows $R_{\mathrm{HK}}^\prime$ as a function of the spectral resolution of the model spectra for two spectra of HD~119850 acquired by the instrument with the lowest resolution in our sample (TIGRE, $\mathcal{R}\approx20\,000$) and the highest resolution (HARPS, $\mathcal{R}\approx115\,000$), both normalized by the $R_{\mathrm{HK}}^\prime$ determined with the respective nominal resolution of the instruments. Both lines converge toward unity around a resolution of 20\,000, leading us to conclude that, while the influence of the spectral resolution is negligible for the instruments in this work, the use of spectra with a lower resolution would result in systematic errors.
   
\begin{figure*}[ht]
\centering
\includegraphics[width=1\hsize]{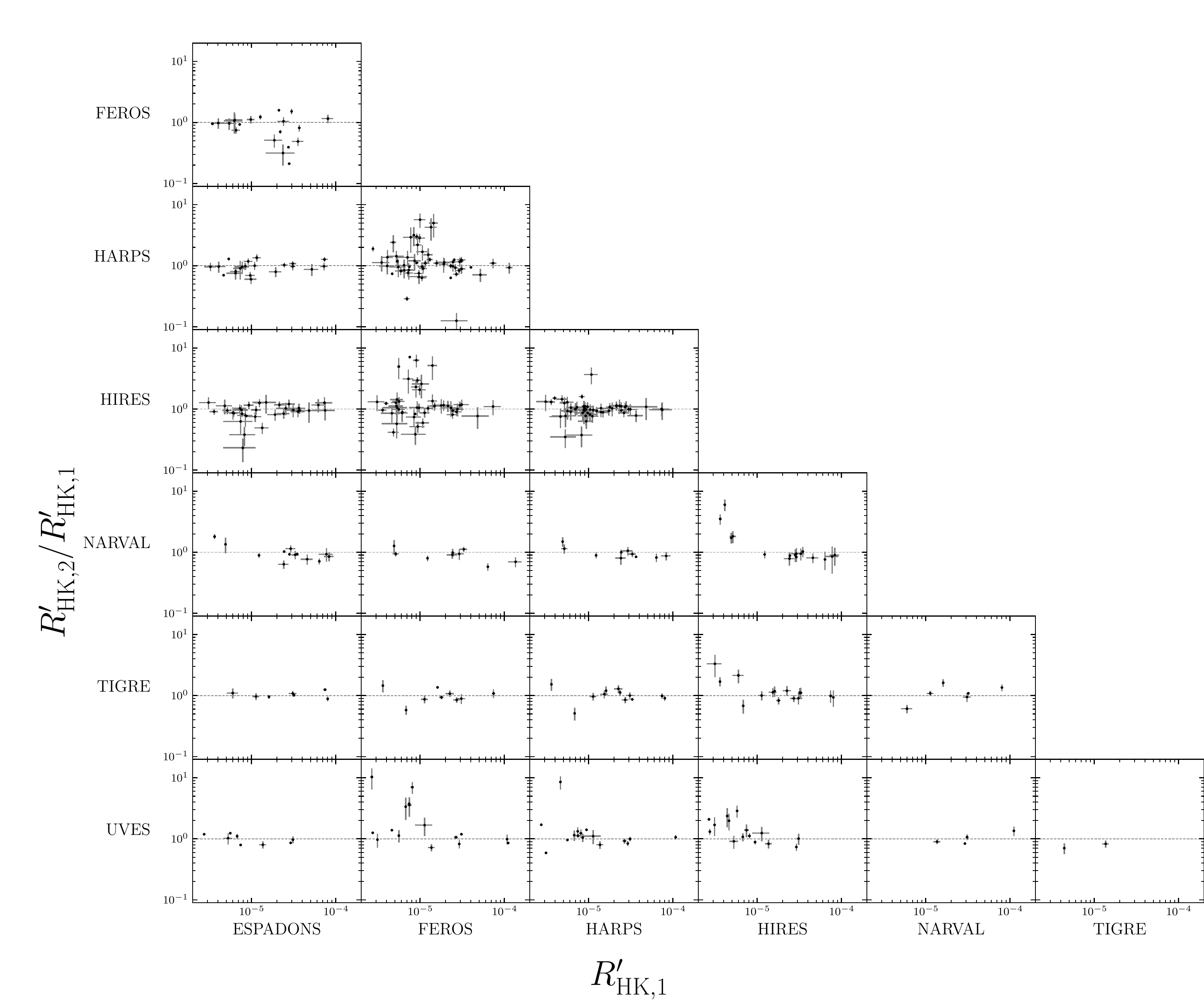}
\captionsetup{width=1\textwidth}
    \caption{Comparison of mean $R_{\mathrm{HK}}^\prime$ derived from individual instruments. The x-axis shows the $R_{\mathrm{HK}}^\prime$ of the instruments denoted by the labels under each subplot, while the y-axis is the fraction between $R_{\mathrm{HK}}^\prime$ of the instruments denoted on the y-axis and those denoted on the x-axis.}
    \label{comparison}
\end{figure*}

We checked the validity of the approach by comparing the time-averaged results from different instruments to each other. 
In order to do this, we evaluated the average $R_{\mathrm{HK}}^\prime$ for each target and for single instruments. 
Figure~\ref{comparison} shows the resulting correlations between the subsamples. 
Apart from a few outliers, all comparisons show a good agreement between data derived from different instruments. We studied the influence of the deviation from equality between values derived from different instruments, and found that it can stem from activity modulation in the form of cycles and/or flares. However, in order to rule out systematic errors, we study the influence of instrumental effects on the accuracy of our method in the next section.

\subsection{Sensitivity to instrumental effects}\label{sec:pol}
\begin{figure}[]
\centering
\includegraphics[width=\hsize]{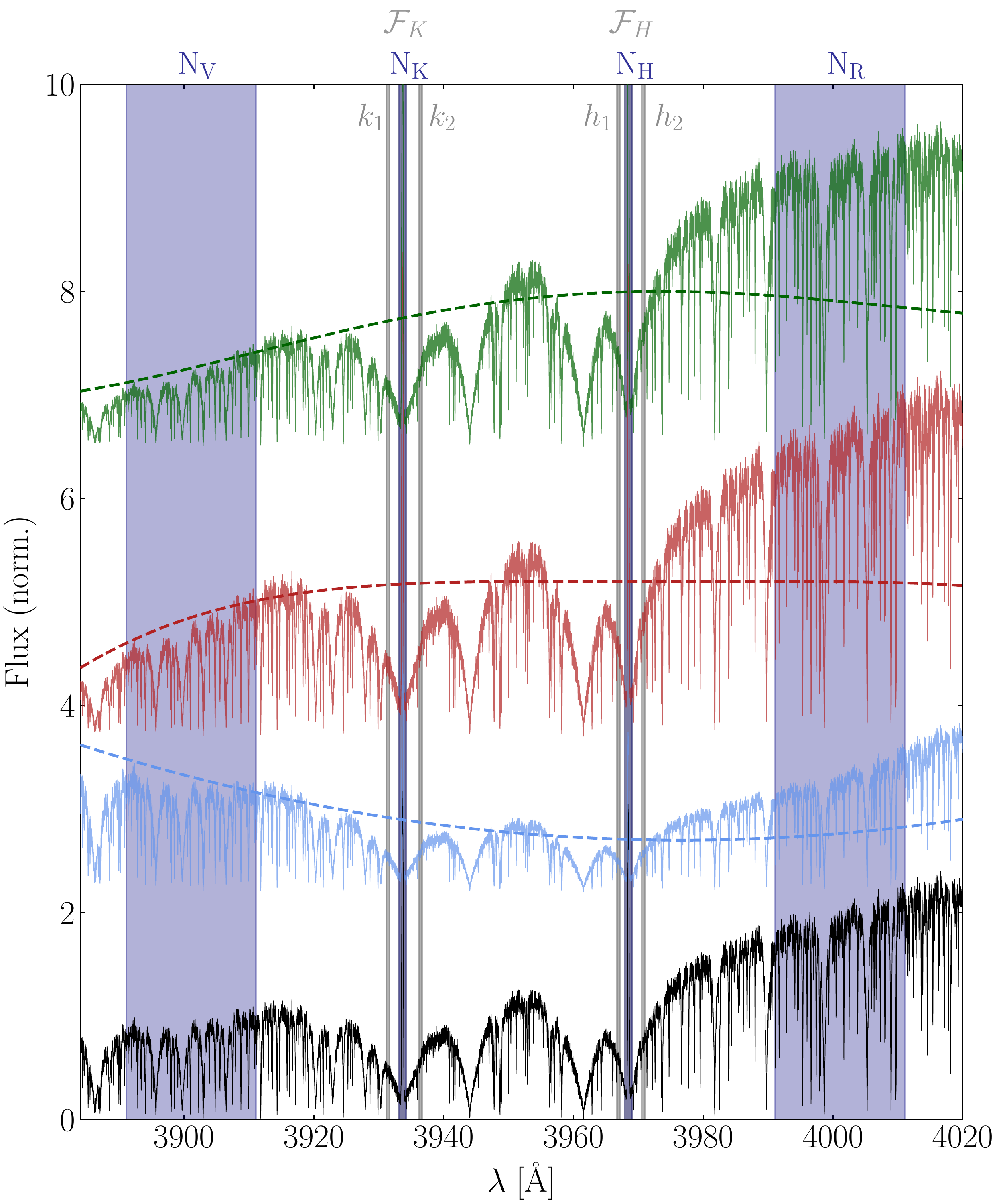}
\captionsetup{width=\hsize}
    \caption{ Example of the artificial instrumental errors. The original HARPS spectrum of Gl 526 (Karmn J13457$+$148) is displayed  at the bottom as a black line, while the polynomials with coefficients stated in Sect.~\ref{sec:pol} and the resulting spectra are shown in blue, red, and green, respectively. The shaded gray areas mark the regions used for the extraction of $R_{\mathrm{HK}}^\prime$ in this publication, whereas the blue regions highlight the bands N$_\mathrm{H}$, N$_\mathrm{K}$, N$_\mathrm{V}$, and N$_\mathrm{R}$ commonly used for the determination of the $S$ index.}
   \label{pol}
\end{figure}

The spectra from different instruments may be subject to systematic errors caused by effects such as the wavelength-dependent throughput, the spectral energy distribution of the flat-fielding source, and the blaze function correction, all of which can cause fluctuations in the normalization. We therefore checked the stability of our approach with regard to variations in the spectral normalization by multiplying a single HARPS spectrum of Gl 526 (displayed in Fig.~\ref{pol}) with a polynomial of shape
\begin{equation}
    p(\lambda)=a_1\left(\frac{\lambda-\lambda_0}{\lambda_{\mathrm{char}}}\right)+a_2\left(\frac{\lambda-\lambda_0}{\lambda_{\mathrm{char}}}\right)^2+a_3\left(\frac{\lambda-\lambda_0}{\lambda_{\mathrm{char}}}\right)^3+a_4\left(\frac{\lambda-\lambda_0}{\lambda_{\mathrm{char}}}\right)^4
\end{equation}
where $\lambda$ is the spectral wavelength, $a_1$, $a_2$, $a_3$, and $a_4$ are polynomial cofficients,  $\lambda_0$ is the central wavelength and $\lambda_{\mathrm{char}}$ is a the characteristic scale of the variability. We repeated the process $10^6$ times, randomly selecting polynomial coefficients $a_{1,2,3,4}$ between $-1$ and $1$, central wavelengths $\lambda_0$ between the cores of the Ca~{\sc ii} H and K lines and $\lambda_{\mathrm{char}}$ between 0.5 and 2 times the average length of the single blaze orders of 51~$\AA$ (HARPS has an effective wavelength range of $3100\,\AA$ and 61 orders). During each try, the relative changes in $R_{\mathrm{HK}}^\prime$ and the classical $S$ index \citep[e.g.,][]{2013A&A...549A.117M, 2017A&A...602A..88A} were evaluated. As an example, Fig.~\ref{pol} shows, aside from the original spectrum, three polynomials and the resulting spectra for coefficients $a_1=[-1,0.5,1]$, $a_2=[1,0.5,-1]$, $a_3=[0,1,-1]$, $a_4=[0,-1,1]$, $\lambda_0=[3951.065\,\AA, 3968.47\,\AA, 3933.66\,\AA]$ and $\lambda_{\mathrm{char}}=[51\,\AA,26\,\AA,102\,\AA]$.

\begin{figure}[]
\centering
\includegraphics[width=\hsize]{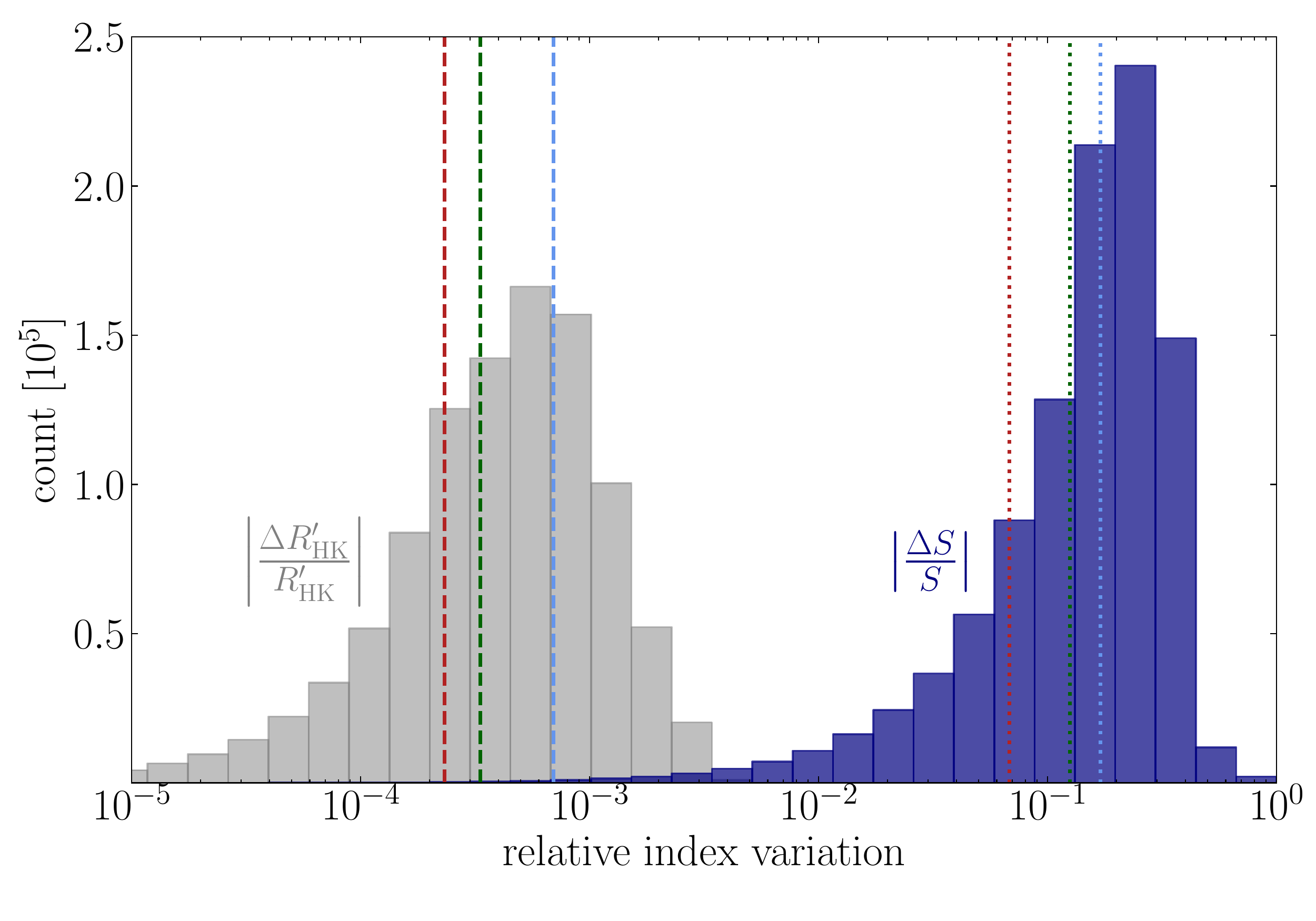}
\captionsetup{width=\hsize}
    \caption{ Distribution of the relative shifts in activity indices. The blue bars represent the $S$ index, whereas the gray bars show $R_{\mathrm{HK}}^\prime$ as determined in this publication. The colored, vertical dashed and dotted lines denote the resulting indices for the three examples shown Fig.~\ref{pol} and discussed in Section~\ref{sec:pol}.}
   \label{pol2}
\end{figure}

Figure~\ref{pol2} shows the resulting distribution of the relative changes $\left| \Delta R_{\mathrm{HK}}^\prime/R_{\mathrm{HK}}^\prime\right|$ and $\left| \Delta S/S \right|$ of the $10^6$ trials, where $R_{\mathrm{HK}}^\prime$ and $S$ are the indices extracted from the unperturbed spectrum, and $\Delta R_{\mathrm{HK}}^\prime$ and $\Delta S$ are the changes relative to them. The peaks of the distributions are separated by more than two orders of magnitude, indicating that our approach is significantly more accurate. We concluded that this is due to the proximity of the reference bands $h_1$, $h_2$, $k_1$, and $k_2$ to the Ca~{\sc ii}~H\&K line cores. Furthermore, the relative errors caused by normalization systematics in our approach are typically of order $10^{-3}$, which is far lower than the typical measurement errors caused by photon noise.

\section{Results and discussion}
\label{sec:4}

\begin{table*}[]
\centering
\small
\caption{Time series of $R_{\mathrm{HK}}^\prime$ and its uncertainty for the last five rows of the table$^{a}$.} \label{table:4}
\begin{tabular}{lcccccc}       
\hline
\hline
\noalign{\smallskip}
Karmn & $\alpha$ & $\delta$ & MJD & $R_{\mathrm{HK}}^\prime$ & $\delta R_{\mathrm{HK}}^\prime$ & Instrument\\ 
~ & [deg] & [deg] & [d] & ~ & ~ & ~ \\ 
\noalign{\smallskip}
\hline
\noalign{\smallskip}
J23505$-$095&357.63183&$-$9.55908&56612.1431395&$1.64\cdot 10^{-5}$&$1.36\cdot 10^{-6}$&FEROS\\
J23505$-$095&357.63183&$-$9.55908&56612.16456367&$1.94\cdot 10^{-5}$&$1.76\cdot 10^{-6}$&FEROS\\
J23505$-$095&357.63183&$-$9.55908&56612.18598737&$1.59\cdot 10^{-5}$&$1.20\cdot 10^{-6}$&FEROS\\
J23492$+$024&357.30221&2.40122&56498.33294243&$3.45\cdot 10^{-6}$&$7.37\cdot 10^{-8}$&FEROS\\
J23492$+$024&357.30221&2.40122&53918.41015954&$6.13\cdot 10^{-6}$&$1.58\cdot 10^{-7}$&HARPS\\
\noalign{\smallskip}
\hline
\end{tabular}
\tablefoot{
\tablefoottext{a}{MJD is the modified Julian date (MJD = JD -- 2\,400\,000.5).
Table 3 is only available in electronic form at the CDS via anonymous ftp to \url{cdsarc.u-strasbg.fr} (130.79.128.5) or via \url{http://cdsweb.u-strasbg.fr/cgi-bin/qcat?J/A+A/}.}
}
\end{table*}

\subsection{Comparison to previous results}
\label{sec:prev}

\begin{figure}[]
\centering
\includegraphics[width=\hsize]{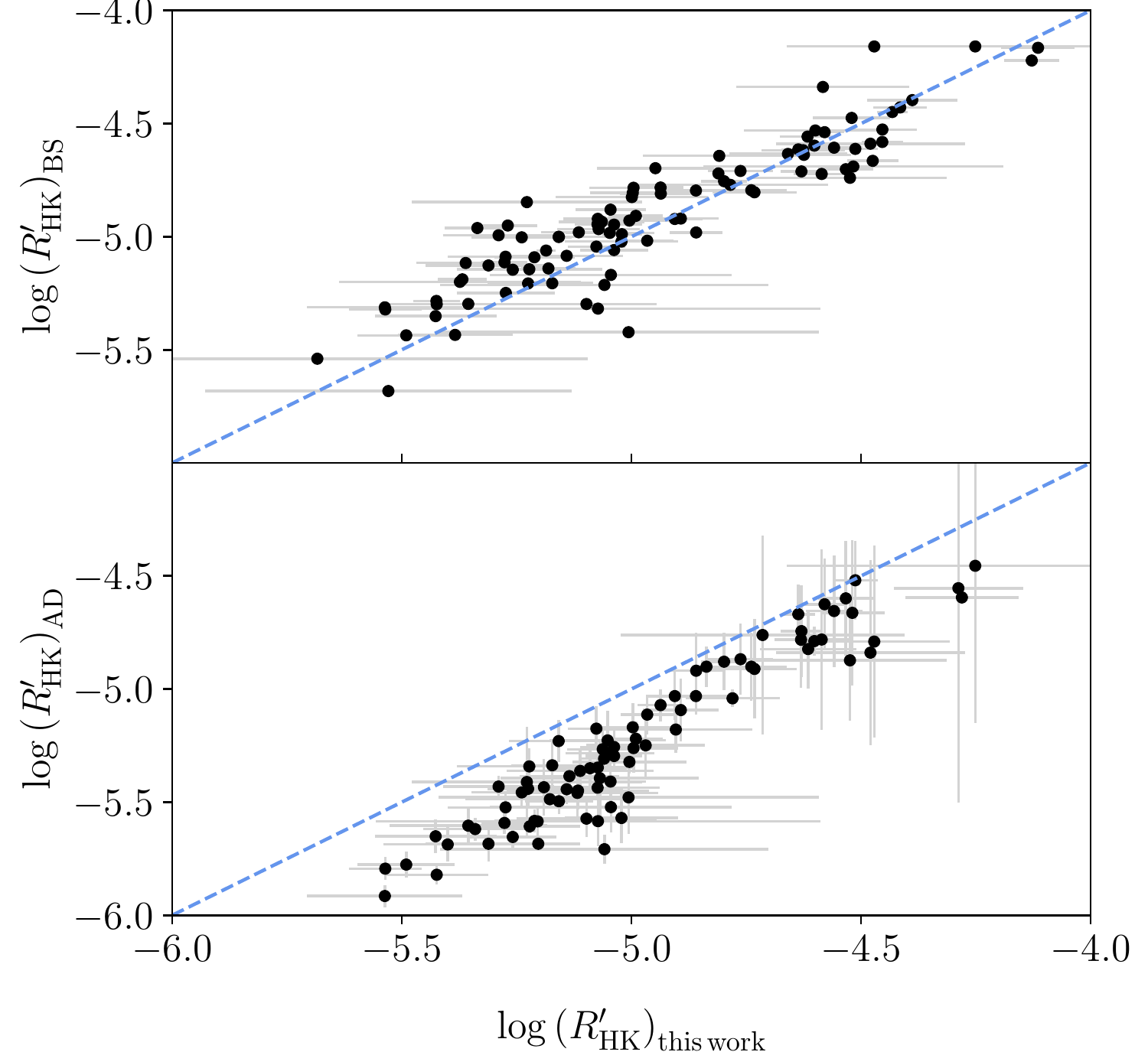}
    \caption{Comparison of $R_{\mathrm{HK}}^\prime$ to catalogs published by \cite{2018A&A...616A.108B} ({\em top panel}) and \cite{2017A&A...600A..13A} ({\em bottom panel}).
    The blue dashed lines indicate equality between the catalogs. 
            }
        \label{activity_comp}
\end{figure}

Several authors have published catalogs of chromospheric Ca~{\sc ii}~H\&K emission of M dwarfs with a sufficiently large overlap with our sample to allow for comparison.
For example, \cite{2017A&A...600A..13A} derived $R_{\mathrm{HK}}^\prime$ of all M dwarfs in the HARPS sample via extraction and conversion of the $S$~index. 
\cite{2018A&A...616A.108B} compiled a catalog of $R_{\mathrm{HK}}^\prime$ of 4454 late-type stars from published data. 
Their measurements were based on the same HARPS spectra as \cite{2017A&A...600A..13A}, which are in turn also included in our database.

Figure~\ref{activity_comp} shows the comparison between our data and those of \cite{2018A&A...616A.108B} and \cite{2017A&A...600A..13A}. While our average values are in good agreement with \cite{2018A&A...616A.108B}, who also used PHOENIX spectra, there is a systematic offset when comparing to \cite{2017A&A...600A..13A} towards lower activity levels. 

\begin{figure}[]
\centering
\includegraphics[width=\hsize]{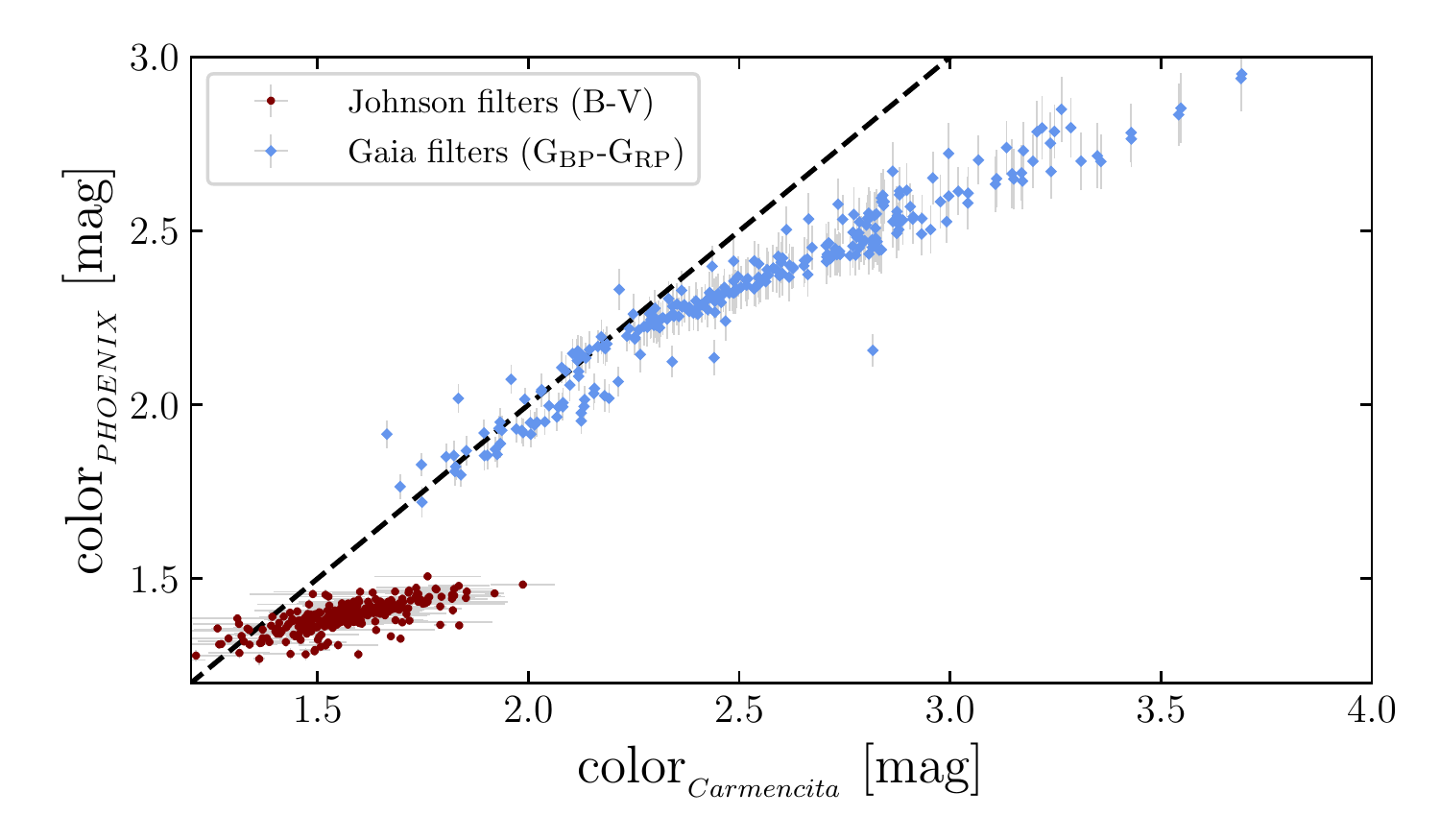}
\captionsetup{width=\hsize}
    \caption{Comparison between measured colors and those determined with PHOENIX synthetic spectra. Johnson $B-V$ and {\em Gaia} $G_{\rm BP}-G_{\rm RP}$ colors are displayed as red dots and blue diamonds, respectively. The dashed line indicates equality between theoretical and measured colors.}
   \label{BVcomp}
\end{figure}

In order to investigate the origin of this discrepancy, we tested the reliability of the flux determination for M dwarfs based on PHOENIX synthetic spectra. In a similar manner as done for the calibration bands in Sect.~\ref{sec:3}, we extracted flux predictions in four filter bands (Johnson $B$ and $V$ and {\em Gaia} $G_{\rm BP}$, $G_{\rm RP}$) from the theoretical spectra. 
Since the magnitudes in these bands have been measured for the majority of the CARMENES GTO sample \citep{2020A&A...642A.115C}, we were then able to compare the measured colors  $G_{BP}-G_{RP}$ and $B-V$ to the theoretical predictions, as displayed in Fig.~\ref{BVcomp}. 
As evident in both colors, the PHOENIX synthetic spectra seem to overestimate the flux in the blue end of the spectrum for the latest M dwarfs.

If true, it explains why we obtained a good agreement with the $R_{\mathrm{HK}}^\prime$ values published by \cite{2018A&A...616A.108B} and a systematic offset compared to those determined by \cite{2017A&A...600A..13A}. 
It also means that any comparison between the activity levels of stars with different parameters will be prone to systematic errors, which is why we did not include the analysis of averaged activity levels of the single targets, for instance, for the derivation of rotation-activity relations, in this publication. 
However, this overestimation does not affect the ability of our method to determine activity time series for individual stars based on data from multiple instruments.

\subsection{Time series}
\label{subsec:2}

\begin{figure*}[]
\centering
\includegraphics[width=\hsize]{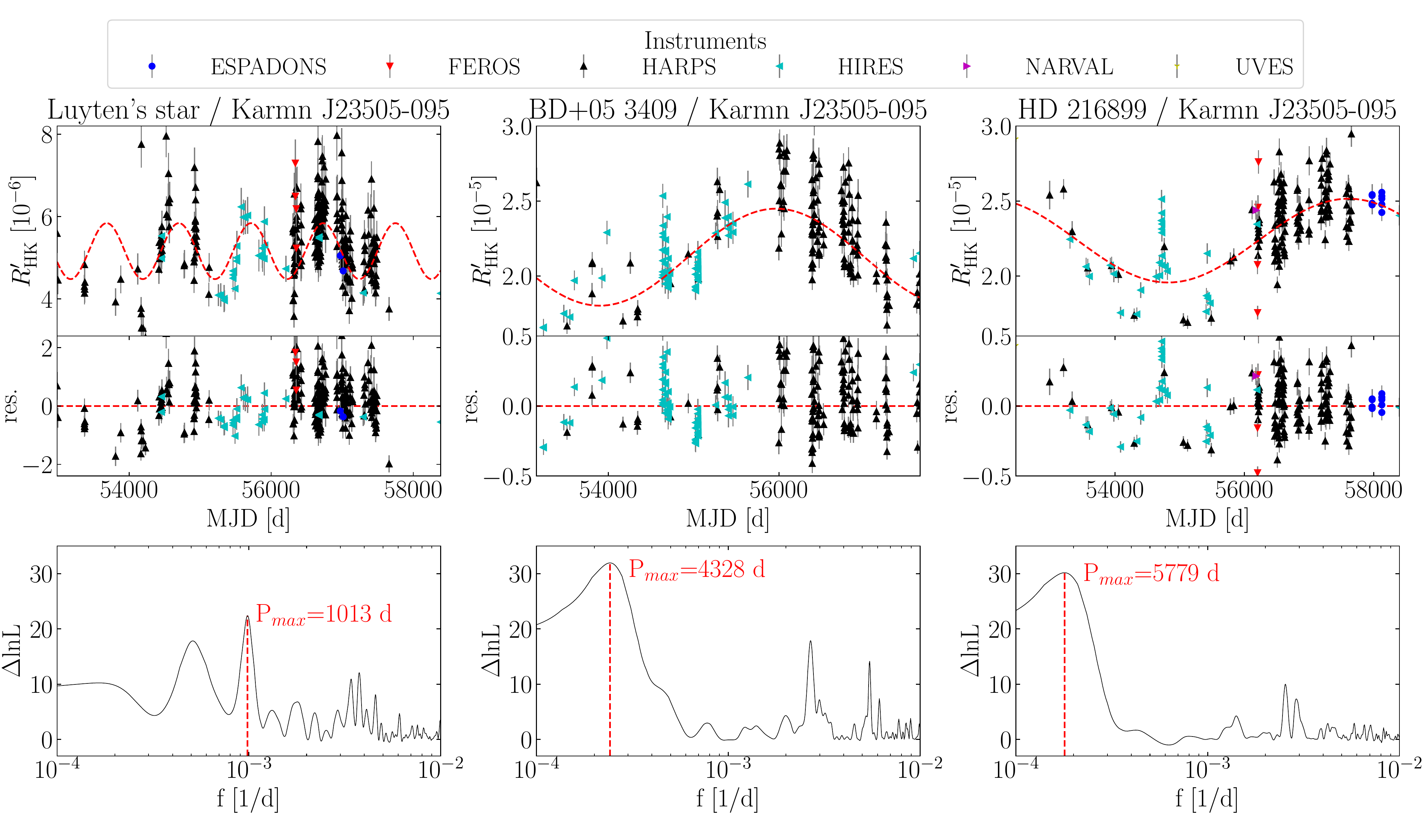}
\captionsetup{width=\hsize}
    \caption{Analysis of the $\rm{R}_{\rm{HK}}^{\prime}$ time series of three M dwarfs exhibiting new long-term cyclic modulation. 
    {\it Upper panels:} $R_{\mathrm{HK}}^\prime$ time series for each star. 
    The color coding and markers denote the instrument (see legend to the right) and the red dashed line is the best-fit sine function. 
    {\it Middle panels:} residuals after subtraction of the best fit. 
    {\it Bottom panels:} GLS 
    periodogram of the time series. 
    The vertical dashed line marks the period with the highest power.}
    \label{fig:c1}
\end{figure*} 

\begin{table}[]
\centering
\small
\caption{New tentative activity cycles of three M dwarfs.} 
\label{table:4}
\begin{tabular}{l c c c}       
\hline
\hline
\noalign{\smallskip}
~ & Star 1 & Star 2 & Star 3 \\
\noalign{\smallskip}
\hline
\noalign{\smallskip}
Karmn & J07274$+$052 & J17303$+$055 & J22565$+$165 \\
Name & \object{Luyten's Star} & \object{BD$+$05 3409} & \object{HD 216899} \\
$P_{\rm rot}^{a}$ [d] & $94 \pm 16$ & ... & $39.5 \pm 0.2$ \\ %
$N_{\rm spectra}$ & $328$ & $196$ & $201$\\ 
\noalign{\bigskip}
$P_{\rm cycle}$ [d] (GLS) & $1012.9\pm4.7$ & $4334\pm78$ & $5663\pm74$ \\
FAP (GLS) & $1.4\,10^{-24}$ & $7.2\,10^{-27}$ & $2.2\,10^{-21}$ \\ 
\noalign{\bigskip}
$P_{\rm cycle}$ [d] & $1012.8\pm4.4$ & $4328\pm90$ & $5780\pm160$ \\
$A$ [$10^{-6}$] & $0.67$ & $3.20$ & $2.78$ \\
$R_{\rm{HK,0}}^{\prime}$ [$10^{-6}$] & $5.18$ & $21.26$ & $22.35$ \\
$\sigma_{\rm jitter}$ & $7.7\,10^{-7}$ & $2.0\,10^{-6}$ & $1.7\,10^{-6}$\\
$\Delta\ln{\mathcal{L}}$ & 21.6 & 34.1 & 28.9 \\ 
\noalign{\smallskip}
\hline
\end{tabular}
\tablefoot{
\tablefoottext{a}{Rotation periods from 
\cite{2017MNRAS.468.4772S}
and \cite{2019A&A...621A.126D}.
}
}

\end{table}

We give below an example of the advantage of activity catalogs derived from multiple instruments. 
The Mount Wilson Observatory HK project \citep[e.g.,][]{1995ApJ...438..269B}, which provides archival $S$~index time series for thousands of stars covering decades, did not include the M dwarf range. 
However, the rise of projects searching for substellar companions to cool dwarfs has resulted in long time series of spectra for M dwarfs, provided the individual data sets are combined.

We present the time series and periodograms of three stars with previously unknown
long-term variations in $R_{\rm HK}^{\prime}$, suggesting the presence of an activity cycle, in Fig.~\ref{fig:c1}.
To estimate the period of this long-term variability, we used the common generalized Lomb-Scargle (GLS) method described by \cite{Zechmeister2009}, along with their approach to estimate the false alarm probability (FAP).
Table~\ref{table:4} lists the resulting periodicities and names of the three stars. 
Since the peaks of all cycles exhibit an FAP smaller than 0.1\,\% and are, thus, significant, we used a sinusoidal model with amplitude $A$, cycle period $P_{\rm cycle}$, offset $R_{\rm{HK,0}}$, and a jitter term $\sigma_{\rm jitter}$ to evaluate the statistical significance of the cyclic variability of the shape. 
The fitted periods, listed in Table~\ref{table:4}, agree well with those derived via GLS. 
We computed the log-likelihood $\ln{\mathcal{L}}$ via:

\begin{equation}
\ln{\mathcal{L}} = -\frac{1}{2}\sum_{i}\,\mathrm{ln}\left( 2\pi (\sigma_i^2+\sigma_{jit}^2) \right)+\frac{\left( R_{\rm{HK,i}}^{\prime}-R_{\rm{HK,mod,i}}^{\prime}\right)^2}{\sigma_i^2+\sigma_{jit}^2}.
\end{equation}

\noindent The best-fit parameters are listed in the bottom panel of Table~\ref{table:4}.

We also implemented a second model that included time-correlated noise following the approach of \cite{2014MNRAS.443.2517H}, thus testing whether the fit quality improves when accounting for rotational variability. While the derived cycle periods agreed with the GLS analysis and the sinusoidal model, and the log-likelihood difference $\Delta\ln{\mathcal{L}}$ increased only slightly compared to the model without red noise, the error estimates did increase significantly, which led us to conclude that such an approach did not constitute a better model.

Two of the three stars had a previous determination of the rotation period, $P_{\rm rot}$ (see Table~\ref{table:4}), which are much shorter than the $P_{\rm cycle}$ that we derive.
\cite{2017A&A...602A..88A} reported a possible $\sim2000$\,d activity cycle of Luyten's Star / Karmn J07274+052, which we identified as an alias of the cycle period in Table~\ref{table:4}.

\section{Summary}
\label{sec:5}

We compiled a library of archival spectra of 186 M dwarfs from the CARMENES GTO sample with sufficiently high S/N acquired by seven high-resolution spectrographs covering the Ca~{\sc ii}~H\&K lines. 
By applying a variant of the spectral subtraction technique and using a set of narrow spectral bands around the doublet lines, we rectified the observed spectra and normalized them based on theoretical PHOENIX model spectra, which enabled us to extract the fluxes in the H\&K lines and, thus, $R_{\mathrm{HK}}^\prime$ in a uniform manner.
This method was shown to be stable with regard to small deviations in radial velocity, and to produce consistent results for data from different instruments.  We also determined that our method is stable with regard to instrumental effects by randomly varying a spectrum with a polynomial. A comparison to the classical $S$ index revealed that our approach is significantly more accurate with respect to instrumental variability.

The aim of this catalog is to provide the basis for further analysis of the $R_{\mathrm{HK}}^\prime$ time series of single targets. 
We demonstrate that the use of a variety of instruments can yield long time series allowing for the discovery of activity cycles.
The detailed derivation of rotation periods and activity cycles of all individual targets is beyond the scope of this paper.

We compared our results to the $R_{\mathrm{HK}}^\prime$ values published by \cite{2018A&A...616A.108B}, who used a similar technique to that employed here, and those derived by \cite{2017A&A...602A..88A}, who converted the ``classical'' $S$~index into $R_{\mathrm{HK}}^\prime$. 
The latter comparison revealed a systematic offset towards low activity levels, the origin of which we determined to be the systematic underestimation of flux in PHOENIX spectra towards the blue end of the spectrum.
While this underestimation prevented us from carrying out a comparison of activity levels of stars with different effective temperatures, it did not influence the purpose of the catalog, namely the creation of long time series of $R_{\mathrm{HK}}^\prime$ for the CARMENES GTO M dwarfs for the determination of rotation periods and other variability.

The catalog is continuously updated when new data become available. Since the approach of using a set of theoretical spectra in conjunction with narrow passbands to rectify and normalize the measured spectrum was proven to be stable, it may be worthwile to apply it to other chromospheric activity indicators.

\begin{acknowledgements}
We thank the anonymous referee for the valuable input.
We acknowledge financial support from the Agencia Estatal de 
Investigaci\'on of the Ministerio de Ciencia, Innovaci\'on y 
Universidades and the ERDF through projects
   PID2019-109522GB-C5[1:4]/AEI/10.13039/501100011033   
and the Centre of Excellence ``Severo Ochoa'' and ``Mar\'ia de 
Maeztu'' awards to the Instituto de Astrof\'isica de Canarias 
(CEX2019-000920-S), Instituto de Astrof\'isica de Andaluc\'ia 
(SEV-2017-0709), and Centro de Astrobiolog\'ia (MDM-2017-0737), and 
the Generalitat de Catalunya/CERCA programme.
Based on observations obtained at the Canada-France-Hawai'i Telescope (CFHT), which is operated by the National Research Council of Canada, the Institut National des Sciences de l´Univers of the Centre National de la Recherche Scientique of France, and the University of Hawai'i. 
Based on observations made with ESO Telescopes at the La Silla Paranal Observatory under programme IDs 0100.C-0097(A), 0101.C-0516(A), 0101.D-0494(A), 0102.C-0558(A), 0102.D-0483(A), 0103.A-9009(A), 072.A-9006(A), 072.C-0488(E), 072.D-0621(A), 073.D-0038(B), 074.B-0639(A), 074.C-0364(A), 074.D-0016(A), 075.D-0614(A), 076.A-9005(A), 076.A-9013(A), 076.C-0155(A), 076.D-0560(A), 077.A-9005(A), 077.C-0364(E), 078.A-9058(A), 078.A-9059(A), 078.C-0044(A), 078.C-0333(A), 078.D-0071(D), 079.A-9007(A), 079.A-9013(B), 079.C-0255(A), 080.D-0086(C), 080.D-0086(D), 080.D-0140(A), 081.A-9005(A), 081.A-9024(A), 081.D-0190(A), 082.C-0218(A), 082.C-0718(A), 082.C-0718(B), 082.D-0953(A), 084.C-0403(A), 085.A-9027(A), 085.C-0019(A), 086.A-9014(A), 087.C-0831(A), 087.C-0991(A), 087.D-0069(A), 088.A-9032(A), 088.C-0662(B), 089.A-9007(D), 089.A-9008(A), 089.C-0440(A), 089.C-0497(A), 089.C-0732(A), 090.A-9003(A), 090.A-9010(A), 090.A-9029(A), 090.C-0200(A), 090.C-0395(A), 091.A-9004(A), 091.A-9012(A), 091.A-9032(A), 091.C-0034(A), 091.C-0216(A), 091.D-0296(A), 092.A-9009(A), 092.C-0203(A), 093.A-9001(A), 093.A-9029(A), 093.C-0343(A), 093.C-0409(A), 094.A-9029(I), 094.D-0596(A), 095.C-0551(A), 095.C-0718(A), 095.D-0685(A), 096.C-0499(A), 097.C-0561(A), 097.C-0561(B), 097.C-0624(A), 097.C-0864(B), 098.C-0518(A), 098.C-0739(A), 099.C-0205(A), 099.C-0880(A), 1102.C-0339(A), 180.C-0886(A), 183.C-0437(A), 183.C-0972(A), 191.C-0505(A), 191.C-0873(A), 191.C-0873(B), 191.C-0873(D), 191.C-0873(E), 191.C-0873(F), 192.C-0224(B), 192.C-0224(C), 192.C-0224(G), 192.C-0224(H), 192.C-0852(A), 192.C-0852(M), 198.C-0838(A), 276.C-5054(A), 60.A-9036(A), 60.A-9709(G), 69.D-0092(A), and 69.D-0478(A).
This research has made use of the Keck Observatory Archive (KOA), which is operated by the W. M. Keck Observatory and the NASA Exoplanet Science Institute (NExScI), under contract with the National Aeronautics and Space Administration.
\end{acknowledgements}

\bibliographystyle{aa}
\bibliography{astro.bib}
\onecolumn
\newpage
\section*{Appendix A: Comparison of $R_{\mathrm{HK}}^\prime$ values from different instruments}

\end{document}